\documentclass{article}
\usepackage{dcase2023,amsmath,graphicx,url,times,booktabs, tabularx}

\usepackage{hhline}
\usepackage{hyperref}
\usepackage{xcolor}
\usepackage{cite}
\usepackage{multirow,url,hyperref}
\usepackage{siunitx}

\usepackage{pifont}
\newcommand{\cmark}{\ding{51}}%
\newcommand{\xmark}{\ding{55}}%

\title{CREATING A GOOD TEACHER FOR KNOWLEDGE DISTILLATION \\ IN ACOUSTIC SCENE CLASSIFICATION}

\name{Tobias Morocutti$^{2}$,
      Florian Schmid$^{1}$,
      Khaled Koutini$^{2}$, 
      Gerhard Widmer$^{1,2}$
      }
\address{$^1$Institute of Computational Perception (CP-JKU), $^2$LIT Artificial Intelligence Lab,\\          
        Johannes Kepler University Linz, Austria \\
        \{tobias.morocutti, florian.schmid, khaled.koutini\}@jku.at\\ 
 }

\begin{document}

\ninept
\maketitle

\begin{sloppy}

\begin{abstract}

Knowledge Distillation (KD) is a widespread technique for compressing the knowledge of large models into more compact and efficient models. KD has proved to be highly effective in building well-performing low-complexity Acoustic Scene Classification (ASC) systems and was used in all the top-ranked submissions to this task of the annual DCASE challenge in the past three years. There is extensive research available on establishing the KD process, designing efficient student models, and forming well-performing teacher ensembles. However, less research has been conducted on investigating which teacher model attributes are beneficial for low-complexity students. In this work, we try to close this gap by studying the effects on the student's performance when using different teacher network architectures, varying the teacher model size, training them with different device generalization methods, and applying different ensembling strategies. The results show that teacher model sizes, device generalization methods, the ensembling strategy and the ensemble size are key factors for a well-performing student network.

\end{abstract}

\begin{keywords}
Acoustic Scene Classification, Knowledge Distillation, CP-Mobile, Patchout FaSt Spectrogram Transformer (PaSST), CP-ResNet
\end{keywords}

\vspace{-2pt}
\section{Introduction}
\label{sec:intro}


The objective of Acoustic Scene Classification (ASC) involves labeling an audio clip with a corresponding scene. The DCASE23 challenge's~\cite{morato22dcase_task} Low-Complexity Acoustic Scene Classification task focuses on utilizing the \textit{TAU Urban Acoustic Scenes 2022 Mobile development dataset (TAU22)}~\cite{heittola20dcase_task}. This dataset comprises one-second audio snippets from ten distinct acoustic scenes. In an attempt to make the models deployable on edge devices, a complexity limit on the models is enforced: models are constrained to have no more than 128,000 parameters and 30 million multiply-accumulate operations (MMACs) for the inference of a 1-second audio snippet. Among other model compression techniques such as Quantization~\cite{hubara17quantized} and Pruning~\cite{frankle2021pruingatinit}, Knowledge Distillation (KD)~\cite{hinton15distilling, Ba14kd, tripathi23kd} proved to be a particularly well-suited technique to improve the performance of a low-complexity model in ASC.

In a standard KD setting, a low-complexity model learns to mimic the teacher by minimizing a weighted sum of hard label loss and distillation loss. While the soft targets are usually obtained by one or multiple possibly complex teacher models, the distillation loss tries to match the student predictions with the computed soft targets based on the Kullback-Leibler divergence.

Jung et al.~\cite{JungKD_ASC2019} demonstrate that soft targets in a teacher-student setup benefit the learning process since one-hot labels do not reflect the blurred decision boundaries between different acoustic scenes.
Knowledge distillation has also been a very popular method in the DCASE challenge submissions. For example, Kim et al.~\cite{kim21tech_report} apply KD using a pretrained teacher. Further, \cite{Lee22tech_report} and \cite{ricardo22tech_report} employ KD to train a low-complexity network on the predictions of a more complex one. 
Schmid et al.~\cite{schmid22KD} use KD to train a low-complexity CNN on a teacher ensemble consisting of five PaSST~\cite{koutini22passt} models. 

To enhance generalization across recording devices, Kim et al. propose a modified version of MixStyle \cite{mixstyle} called Freq-MixStyle \cite{kim22rfn, schmid22KD}. This method normalizes each frequency band and denormalizes it with mixed frequency statistics of two different samples. 

Another method for improving the device generalization is Device Impulse Response Augmentation \cite{morocutti23dir} which was introduced by Morocutti et al. It convolves audio signals with impulse responses of vintage microphones to increase the recording device variety in the training phase.

In this work, we study the effects of training a low-complexity network on the predictions of a single teacher or a teacher ensemble. We experiment with different network architectures, model sizes and device generalization methods to create the single teacher model that leads a student to perform best on the validation set. 
Additionally, we analyze the effect of combining teacher models with different network architectures, sizes, or device generalization methods. 

\vspace{-2pt}
\section{Network Architectures}
\label{sec:network_arch}

We experiment with three different teacher networks that were shown~\cite{Schmid23tech_report} to perform well as a teacher for the task of ASC.
The architectures consist of two receptive-field regularized~\cite{koutini21journal} convolutional neural networks (CNNs):  CP-ResNet~\cite{Koutini2019Receptive} and CP-Mobile~\cite{Schmid23tech_report}, as well as a Transformer model:  Patchout faSt Spectrogram Transformer (PaSST)~\cite{koutini22passt}.

\vspace{-5pt}
\subsection{CP-Mobile}

CP-Mobile (CPM)~\cite{Schmid23tech_report} is an efficient architecture optimized for ASC. This architecture is designed to be less complex than CP-ResNet by factorizing convolution operations, such as in MobileNets~\cite{Sandler18MobileNetsV2, Howard19MobileNetV3} and EfficientNets~\cite{Tan19EfficientNet}, while maintaining important properties that were shown to be important for ASC tasks, such as the regularized receptive field~\cite{koutini21journal, Koutini2019Receptive}.

In the following experiments, the student model has the CPM architecture with the following configuration: 32 base channels, an expansion rate of 3 and a channels multiplier of 2.3. 
The details of the CPM architecture are described in~\cite{Schmid23tech_report}. In short, these attributes control the scale of the network: the base channels represent the width of the first few blocks of the network; the channels multiplier determines the expansion in the number of channels as the network gets deeper (i.e. the number of channels in the last convolutional blocks is the number of channels of the previous blocks multiplied by channels multiplier); the expansion rate determines the number of channels in the depthwise convolution. The resulting model consists of almost 128K parameters and 29 million multiply-accumulate operations (MMACs). 

We choose CPM as a student model since the architecture is designed for low-complexity ASC and has been shown to outperform CP-ResNet in previous work~\cite{Schmid23tech_report}.
In addition, we experiment with using a scaled-up version of CPM as a teacher model for KD. To scale up the network, we increase the width via the base-channels hyperparameter.

\vspace{-5pt}
\subsection{CP-ResNet}

CP-ResNet (CPR)~\cite{Koutini2019Receptive, koutini21journal} is a receptive-field regularized CNN which has been shown to be very successful for ASC in previous editions of the DCASE ASC challenge~\cite{morato22dcase_task, heittola20dcase_task, koutini21dcasesubmission, koutini20dcasesubmission}. Therefore, we also use this network as a teacher model. We use the number of base channels to scale up the network in order to create teacher models with different sizes, similar to the procedure outlined for CPM.

\vspace{-5pt}
\subsection{PaSST}

The Patchout faSt Spectrogram Transformer (PaSST)~\cite{koutini22passt} is a complex, self-attention-based model, which is pre-trained on AudioSet~\cite{gemmeke17audioset} and consists of 85M parameters. The pre-trained model can be fine-tuned to achieve state-of-the-art performances on multiple downstream tasks, including ASC~\cite{koutini22passt}. Additionally, PaSST models have proven to be excellent teachers for low-complexity CNNs~\cite{schmid22KD, schmid22cpjku, schmid23efficient}. Therefore, we also experiment with PaSST as a teacher model.

\vspace{-2pt}
\section{Knowledge Distillation}
\label{sec:kd}

We train our student model on the pre-computed predictions of the teacher model or teacher ensemble in addition to the one-hot encoded labels, similar to~\cite{schmid23efficient}.
Training the student model on the soft labels of the teacher (ensemble) results in the student model learning blurred decision boundaries and establishing important similarity structures between classes. The loss is given in Equation \ref{eq:ts_loss} and consists of the hard label loss $L_t$ and distillation loss $L_{kd}$. The label and distillation loss are weighted using the factor $\lambda$. The student and teacher logits are denoted by $z_s$ and $z_t$, while $y$ stands for the hard labels. $\tau$ is a temperature to control the sharpness of the probability distributions created by the softmax activation $\delta$. $L_l$ indicates the Cross-Entropy loss and the Kullback Leibler divergence is used as distillation loss $L_{kd}$. 

\vspace{-10pt}
\begin{equation}
  \label{eq:ts_loss}
    Loss = \lambda L_l(\delta(z_S), y) + (1 - \lambda) \tau^2 L_{kd}(\delta(z_S/\tau), \delta(z_T/\tau))
\vspace{-2pt}
\end{equation}

As suggested in \cite{hinton15distilling}, we multiply the distillation loss by $\tau^2$ since the magnitudes of the gradients produced by the soft targets scale as 1/${\tau^2}$. This ensures that the relative contributions of the hard and soft targets remain roughly unchanged if the temperature used for distillation is modified.

\vspace{-5pt}
\subsection{Experimental Setup}
\label{sec:exp_setup}

We train the teacher models as well as the student models on the TAU22~\cite{heittola20dcase_task} dataset with the shifted crops dataset augmentation described in \cite{Schmid23tech_report}. Regarding Knowledge Distillation, we use the values of 0.02 and 2 for $\lambda$ and temperature $\tau$, respectively.

For device generalization (DG) we experiment with Freq-MixStyle (FMS)~\cite{kim22rfn, schmid22KD} and Device Impulse Response (DIR) augmentation~\cite{morocutti23dir} and the combination thereof. FMS is configured by two parameters: $\alpha_{fms}$ determines the shape of the Beta distribution used to randomly draw mixing coefficients, and $p_{fms}$ specifies the probability of whether it is applied to a batch or not. Similar to FMS, DIR is guided by a probability $p_{dir}$ that determines the augmentation strength by specifying the proportion of samples to augment. 

The configurations used for FMS and DIR are adapted for each architecture. Results in~\cite{morocutti23dir} show that PaSST performs best with $\alpha_{fms}=0.4$, $p_{fms}=0.4$ and $p_{dir}=0.6$ whereas CPR achieves the highest validation accuracy using $\alpha_{fms}=0.3$, $p_{fms}=0.8$ and $p_{dir}=0.4$. While our experiments found that CPM teachers perform well using the same configuration as used for CPR, setting $\alpha_{fms}$, $p_{fms}$ and $p_{dir}$ to 0.3, 0.4 and 0.6 when training the student network results in higher validation accuracy. More details about our experimental setup are reported in \cite{Schmid23tech_report}.

\vspace{-2pt}
\section{Single Model Teacher}
\label{sec:single_model_teacher}

In this section, we compare the performance of different teachers and evaluate the performance of students trained on the predictions of different teacher models using KD. We experiment with using a single CPM, CPR or PaSST model as the teacher and a low-complexity CPM as the student.

\vspace{-5pt}
\subsection{Scaling the Teacher}
\label{sec:compare_single_arch}

To investigate the effect of training the student on teachers of different complexity, we scale CPM and CPR by increasing the number of base channels, which modifies the width of the network. We test the effect of scaling the teacher only on CPM and CPR since we use a pre-trained PaSST model.

We trained CPM and CPR models in five different complexity configurations such that their number of parameters is approximately 128K, 450K, 1M, 4M and 8M. Since the number of parameters of CPM and CPR does not scale equally when increasing the base channels, we selected the number of base channels for each size and architecture individually. We used 32, 56, 88, 168 and 232 base channels for CPR and 32, 64, 96, 184 and 264 base channels for CPM. 

All different configurations are evaluated over three runs and to ensure that our experiments are independent of each other, we train one student on each of the three teachers.

Additionally, we apply a combination of Freq-MixStyle and Device Impulse Response augmentation to all student as well as all teacher models. From now on, we will refer to the combination of DIR and FMS as DIRFMS. 

\renewcommand{\arraystretch}{1.1}
\begin{table}[h]
\vspace{-0pt}
\begin{center}
\begin{tabular}{c|l|c|c|c|c|c|c}
\toprule
 \multicolumn{2}{c|}{} & \multicolumn{2}{c|}{\textbf{CPR}} & \multicolumn{2}{c|}{\textbf{CPM}} & \multicolumn{2}{c}{\textbf{PaSST}}  \\ 
 \multicolumn{2}{c|}{} & T & S & T & S & T & S \\ \midrule
\parbox[t]{2mm}{\multirow{6}{*}{\rotatebox[origin=c]{90}{\textbf{Teacher size}}}} & \textbf{128K}
& 60.28 & 63.94 & 62.66 & \textbf{63.70} & - & - \\ 

& \textbf{450K} 
& 62.05 & \textbf{64.60} & 62.81 & 62.48 & - & - \\ 

& \textbf{1M} 
& 62.58 & 63.99 & 63.92 & 62.76 & - & - \\

& \textbf{4M}
& 62.74 & 63.51 & 64.28 & 62.43 & - & - \\

& \textbf{8M}
& \textbf{63.28} & 63.43 & \textbf{64.62} & 62.83 & - & - \\

& \textbf{85M}
& - & - & - & - & \textbf{62.20} & \textbf{64.65} \\

 \bottomrule

\end{tabular}
\caption{Validation accuracy of different teacher networks, and a student model trained on these. T and S denote the performance of the teacher and student, respectively. While the teacher networks vary in architecture and size, the student model is always a CPM model with 128k parameters. All results are averages over three independent runs and the last 4 epochs of training.}
\label{tab:scaling_teacher}
\end{center}
\vspace{-12pt}
\end{table}
\renewcommand{\arraystretch}{1}

Table \ref{tab:scaling_teacher} shows that for the teacher, CPM outperforms CPR in each complexity configuration. Additionally, even the smallest variant of CPM achieves a higher validation accuracy than PaSST, which has several orders of magnitude more parameters. 

However, the students trained on CPM perform worse than the ones trained on CPR for each size of teacher. Furthermore, the students trained using PaSST as a teacher outperform the best students trained on a CPR variant by only 0.05\%.
While the teacher with 450K parameters works best for CPR, the variant with 128K parameters makes the best CPM teacher. 

In short, the results show that the right scale of a CNN teacher can improve the performance of the students by more than 1\%. Furthermore, smaller CNNs can be better teachers, even if the larger teachers outperform the smaller ones. Finally, having a different architecture for teacher and student improves the performance of the student. 

\vspace{-5pt}
\subsection{Effect of Device Generalization Methods}
\label{sec:effect_of_single_augmentation}

Table \ref{tab:device_generalization_methods} presents the impact of the device generalization (DG) methods DIR, FMS and DIRFMS.
For studying the effects of these methods, we use the teacher variations with 128K and 450K parameters for CPM and CPR, respectively, since these teacher models result in the best performing student models, as shown in Section~\ref{sec:compare_single_arch}.

\renewcommand{\arraystretch}{1.1}
\begin{table}[h]
\vspace{-5pt}
\begin{center}
\begin{tabular}{l|c|c|c|c|c|c}
\toprule
 & \multicolumn{2}{c|}{\textbf{CPR}} & \multicolumn{2}{c|}{\textbf{CPM}} & \multicolumn{2}{c}{\textbf{PaSST}}  \\ 
 & T & S & T & S & T & S \\ \midrule \midrule
 \multicolumn{7}{c}{\textbf{Validation Accuracy}} \\ \midrule
 \textbf{DIRFMS}
& \textbf{62.05} & \textbf{64.60} & \textbf{62.66} & \textbf{63.70} & \textbf{62.20} & \textbf{64.65} \\ 

\textbf{DIR} 
& 57.34 & 62.47 & 57.23 & 61.57 & 61.64 & 64.39 \\

\textbf{FMS} 
& 60.99 & 63.40 & 61.18 & 63.66 & 61.08 & 64.56 \\ 

\textbf{NO AUG}
& 54.13 & 62.74 & 53.15 & 62.47 & 59.39 & 63.76 \\ \midrule \midrule

 \multicolumn{7}{c}{\textbf{Unseen Accuracy}} \\ \midrule
 \textbf{DIRFMS}
& \textbf{56.95} & \textbf{60.43} & \textbf{57.92} & \textbf{59.20} & \textbf{58.73} & \textbf{61.03} \\

\textbf{DIR} 
& 49.30 & 56.74 & 48.62 & 55.54 & 57.91 & 60.90 \\

\textbf{FMS} 
& 54.94 & 58.91 & 54.92 & 58.76 & 57.57 & 61.00 \\

\textbf{NO AUG}
& 44.75 & 56.70 & 43.94 & 56.21 & 54.08 & 59.60 \\
\bottomrule

\end{tabular}
\caption{Validation accuracy of teacher networks trained using different DG methods, and a student model trained on the corresponding teacher predictions. T and S denote the performance of the teacher and student, respectively. The CPM teacher has 128K parameters, the CPR teacher has 450K parameters. While the teacher network varies in architecture and used DG method, the student is always a CPM model with 128k parameters trained with DIRFMS. All results are averages over three independent runs and the last 4 epochs of training.}
\label{tab:device_generalization_methods}
\end{center}
\vspace{-12pt}
\end{table}
\renewcommand{\arraystretch}{1}

The results in Table \ref{tab:device_generalization_methods} show that FMS, DIR and/or DIRFMS boost both the performance of the teacher models as well as the performance of the student models significantly.
The results show that there is a clear effect of these methods on the validation accuracy. Moreover, this effect tends to be even higher on the unseen accuracy.
Applying DIRFMS results in the best validation and unseen accuracy, outperforming DIR and FMS. We define \textit{unseen accuracy} as the accuracy on the subset of the validation set that consists of samples of devices not present in the training set. Consistent with the findings in~\cite{morocutti23dir}, FMS, DIR and DIRFMS have less effect on the performance of PaSST, compared to CPR or CPM. 

\vspace{-2pt}
\section{Ensemble Teacher}
\label{sec:ensemble_teacher}

Previous work~\cite{Schmid23tech_report} shows that training the student on the predictions of multiple teacher networks is a highly effective method to improve the accuracy of the student in the KD framework. This effect is even more significant when ensembling different architectures or models trained with different device generalization methods.
In this section, we will experiment with different ensemble configurations and show their effect on the low-complexity student.
We ensemble different models by averaging their logits.

\renewcommand{\arraystretch}{1.1}
\begin{table}[h]

\begin{center}
\begin{tabular}{l|c|c|c|c|c}
\toprule

& \multicolumn{2}{c|}{\textbf{CPR}} & \multicolumn{2}{c|}{\textbf{CPM}} & \textbf{PaSST}  \\ \midrule

size of teacher & 128K & 450K & 128K & 450K & 85M  \\ \midrule
 
 \textbf{1 teacher}
& 63.94 & \textbf{64.60} & \textbf{63.70} & 62.48 & \textbf{64.65} \\ 

\textbf{3 teacher} 
& \textbf{64.53} & 64.36 & \textbf{63.97} & 62.77 & \textbf{64.81} \\ 

 \bottomrule
\end{tabular}
\caption{Validation accuracy of student models trained on the predictions of either one or three teacher models which apply both Freq-MixStyle and Device Impulse Response augmentation (\textbf{DIRFMS}). The highest accuracy per architecture and per number of teacher is marked bold. For CPR and CPM, the teacher models consist of either 128K or 450K parameters. All results are averages over three independent runs and the last 4 epochs of training.}
\label{tab:same_teacher}
\end{center}
\vspace{-12pt}
\end{table}
\renewcommand{\arraystretch}{1}

\renewcommand{\arraystretch}{1.1}
\begin{table*}[h!]
\vspace{-5pt}
\begin{center}
\begin{tabular}{l|c|c|c||c|c|c|c}
\toprule
\multicolumn{8}{c}{\textbf{Teacher Ensemble Variations}} \\
\midrule \midrule

\multicolumn{8}{l}{\textbf{Teacher Architecture}} \\
\midrule

CPR & \cmark & \xmark & \xmark & \cmark & \cmark & \xmark & \cmark \\
CPM & \xmark & \cmark & \xmark & \xmark & \cmark & \cmark & \cmark \\
PaSST & \xmark & \xmark & \cmark & \cmark & \cmark & \cmark & \xmark \\
\midrule \midrule 
\multicolumn{8}{l}{\textbf{Device Generalization Methods}} \\ \midrule

DIR + FMS 
& 64.25 & 62.35 & 64.47
& - & - & - & - \\

DIRFMS + DIR 
& 64.21 & 63.45 & 64.63 
& - & - & - & - \\ \midrule

DIRFMS
& 64.53 & \textbf{63.97} & 64.81
& 65.19 & 65.09 & \textbf{65.15} & 64.66 \\ 

DIRFMS + FMS
& \textbf{64.74} & 63.76 & 64.89
& \textbf{65.81} & 65.12 & 64.67 & \textbf{64.67} \\  

DIRFMS + DIR + FMS
& 64.10 & 63.76 & \textbf{65.16}
& 65.39 & \textbf{65.18} & 64.85 & 64.03 \\  
\bottomrule
\end{tabular}
\caption{The accuracy of the student model being trained on a teacher ensemble. The teacher ensembles differ in the combination of architectures and the combination of DG methods. A mark indicates that three models of the corresponding architecture are included in the ensemble. All results are averages over three independent runs and the last 4 epochs of training.}
\label{tab:multiple_archs}
\end{center}
\vspace{-12pt}
\end{table*}
\renewcommand{\arraystretch}{1}

\vspace{-5pt}
\subsection{Ensembling Teachers with Identical Training Setup}
\label{sec:combine_diff_seed}

This section presents experiments about ensembling different models that use the same training setup but different seeds. More precisely, we ensemble different models that share the same architecture, complexity and DG methods. The goal is to test if the averaged logits of multiple teacher models are better soft targets for training the student model.

Since the results in Table \ref{tab:device_generalization_methods} indicate that DIRFMS has the most positive effect on the students for all teacher architectures, we evaluate the performance of students learning from a teacher ensemble trained with DIRFMS. Additionally, we choose to test the training of the student on the teacher ensembles with two different complexity configurations of the CPR and CPM teachers. Due to the fact that CPR performs best with 450K and CPM with 128K parameters, we select these two complexity levels to evaluate the teacher ensembling on both architectures.

As the results in Table \ref{tab:same_teacher} show, the CPR teacher with 128K parameters outperforms the variant with 450K parameters when using an ensemble of three teachers. Further, the variant with 128K parameters also works best for the CPM teacher, outperforming the 450K-parameters variant by 1.2\%. When we train the students on the averaged logits of three PaSST models, the validation accuracy of the student increases slightly by 0.16\%, compared to using only one PaSST teacher. However, PaSST outperforms the other architectures, with CPM performing worse than CPR.

\vspace{-5pt}
\subsection{Ensembling Teachers Trained with Different DG Methods}
\label{sec:combine_augmentation_methods}

In this section, we experiment with combining models with the same architecture but trained using different DG methods in order to create a better teacher ensemble. We choose 128K parameters as the teacher complexity for CPR and CPM, since this complexity performs best when combining multiple models, as shown in Table \ref{tab:same_teacher}. We evaluate the effect of training the student on these teacher ensembles and compare the results with the performance of the students trained using the DIRFMS teacher ensemble described in Section \ref{sec:combine_diff_seed}. All evaluated teacher ensembles contain three models for each included DG method. This implies that the different ensembles stated in the left part of Table \ref{tab:multiple_archs} contain between 3 and 9 models. 

The results in Table \ref{tab:multiple_archs} indicate that including teachers trained using DIRFMS in the ensemble is essential for every architecture, since the ensembles DIRFMS+FMS, DIRFMS and DIRFMS+DIR+FMS perform best for the CPR, CPM and PaSST architecture, respectively. Including the DIR teacher in the DIRFMS+FMS ensemble only increases the performance of students trained on the predictions of PaSST models. The best-evaluated ensemble of only one architecture is the PaSST DIRFMS+DIR+FMS ensemble, increasing the accuracy by 0.35\% compared to the previously best PaSST DIRFMS ensemble. 

\vspace{-5pt}
\subsection{Ensembling Teachers with Different Architectures}
\label{sec:combine_archs}

In this section, we experiment with ensembling different architectures motivated by the assumption that different architectures can learn different features and aspects of the training data and therefore ensembling them would result in a more robust model.

We test each combination of CPR, CPM and PaSST using the combinations of DG methods, which performed best on single architecture ensembles. 
It is worth noting that the teacher ensemble size depends on the number of used architectures and DG methods. It can therefore range from 6 (2 architectures x 1 DG method x 3 models) to 27 (3 architectures x 3 DG methods x 3 models). 

The results in Table \ref{tab:multiple_archs} clearly show that the teacher ensembles consisting of CPR and PaSST models result in the best-performing students. Adding CPM models to ensembles of CPR and PaSST models worsens the performance of the students for all evaluated DG configurations. 
More precisely, ensembling CPM and CPR does not lead to performance improvement, and neither does ensembling CPM and PaSST.

Regarding the DG methods, ensembling teacher models trained with DIRFMS and FMS results in the best student performance for the CPR and PaSST combination, creating the best-evaluated ensemble with 65.81\% validation accuracy of the student.  

\vspace{-2pt}
\section{Conclusion}
\label{sec:conclusion}

In this work, we show that low-complexity CNNs like the CPM learn more important features from Transformers or relatively small CNNs compared to large CNNs when using Knowledge Distillation. 
Additionally, we show that applying Device Impulse Response (DIR) augmentation, Freq-Mixstyle (FMS) and especially the combination thereof (DIRFMS) to the teacher models significantly boosts the performance of the teachers and the students. The effect of these DG methods is even more noticeable on the unseen accuracy, compared to the total validation accuracy.
Surprisingly, it turns out that the performance of the student does not necessarily improve with the scale of the teacher. For example, ensembling smaller teacher networks can be more beneficial than ensembling bigger ones.
Furthermore, we show that the performance of the student improves when the teacher architecture is different than the student architecture. For example, when using PaSST or CPR to train CPM. In contrast, the low-complexity CPM student performs worse when it is trained on any higher complexity variation of the same architecture. Additionally, the predictions of PaSST and CPR complement each other, resulting in better student performance. 
Finally, using an ensemble of CPR and PaSST trained either using DIRFMS or FMS results in our best student, which has an accuracy of 65.81\% with 128K parameters and 32 million MACCS, outperforming the much larger CPR, CPM and PaSST models.  

\vspace{-2pt}
\section{Acknowledgment}
\label{sec:acknowledgment}

The LIT AI Lab is supported by the Federal State of Upper Austria. GW's work is supported by the European Research Council (ERC) under the European Union's Horizon 2020 research and innovation programme, grant agreement No 101019375 (Whither Music?).


\bibliographystyle{IEEEtran}
\bibliography{refs}

\begin{thebibliography}{10}
\providecommand{\url}[1]{#1}
\def\UrlFont{\rmfamily}
\providecommand{\newblock}{\relax}
\providecommand{\bibinfo}[2]{#2}
\providecommand\BIBentrySTDinterwordspacing{\spaceskip=0pt\relax}
\providecommand\BIBentryALTinterwordstretchfactor{4}
\providecommand\BIBentryALTinterwordspacing{\spaceskip=\fontdimen2\font plus
\BIBentryALTinterwordstretchfactor\fontdimen3\font minus
  \fontdimen4\font\relax}
\providecommand\BIBforeignlanguage[2]{{%
\expandafter\ifx\csname l@#1\endcsname\relax
\typeout{** WARNING: IEEEtran.bst: No hyphenation pattern has been}%
\typeout{** loaded for the language `#1'. Using the pattern for}%
\typeout{** the default language instead.}%
\else
\language=\csname l@#1\endcsname
\fi
#2}}

\bibitem{morato22dcase_task}
I.~Mart\'{i}n-Morat\'{o}, F.~Paissan, A.~Ancilotto, T.~Heittola, A.~Mesaros,
  E.~Farella, A.~Brutti, and T.~Virtanen, ``Low-complexity acoustic scene
  classification in {DCASE} 2022 challenge,'' in \emph{DCASE Workshop}, 2022.

\bibitem{heittola20dcase_task}
T.~Heittola, A.~Mesaros, and T.~Virtanen, ``Acoustic scene classification in
  {DCASE} 2020 challenge: Generalization across devices and low complexity
  solutions,'' in \emph{DCASE Workshop}, 2020.

\bibitem{hubara17quantized}
I.~Hubara, M.~Courbariaux, D.~Soudry, R.~El{-}Yaniv, and Y.~Bengio, ``Quantized
  neural networks: Training neural networks with low precision weights and
  activations,'' \emph{J. Mach. Learn. Res.}, 2017.

\bibitem{frankle2021pruingatinit}
J.~Frankle, G.~K. Dziugaite, D.~Roy, and M.~Carbin, ``Pruning neural networks
  at initialization: Why are we missing the mark?'' in \emph{{ICLR}}, 2021.

\bibitem{hinton15distilling}
G.~E. Hinton, O.~Vinyals, and J.~Dean, ``Distilling the knowledge in a neural
  network,'' \emph{CoRR}, 2015.

\bibitem{Ba14kd}
J.~Ba and R.~Caruana, ``Do deep nets really need to be deep?'' in
  \emph{NeurIPS}, 2014.

\bibitem{tripathi23kd}
A.~M. Tripathi and O.~J. Pandey, ``Divide and distill: New outlooks on
  knowledge distillation for environmental sound classification,'' \emph{{IEEE}
  {ACM} Trans. Audio Speech Lang. Process.}, 2023.

\bibitem{JungKD_ASC2019}
H.~Heo, J.~Jung, H.~Shim, and H.~Yu, ``Acoustic scene classification using
  teacher-student learning with soft-labels,'' in \emph{{Interspeech}}.\hskip
  1em plus 0.5em minus 0.4em\relax {ISCA}, 2019.

\bibitem{kim21tech_report}
B.~Kim, S.~Yang, J.~Kim, and S.~Chang, ``{QTI} submission to {DCASE} 2021:
  Residual normalization for device-imbalanced acoustic scene classification
  with efficient design,'' DCASE2021 Challenge, Tech. Rep., 2021.

\bibitem{Lee22tech_report}
J.-H. Lee, J.-H. Choi, P.~M. Byun, and J.-H. Chang, ``Hyu submission for the
  {DCASE} 2022: Efficient fine-tuning method using device-aware
  data-random-drop for device-imbalanced acoustic scene classification,''
  DCASE2022 Challenge, Tech. Rep., 2022.

\bibitem{ricardo22tech_report}
R.~Anastácio, L.~Ferreira, F.~Mónica, and C.~B. Luís, ``Ai4edgept submission
  to {DCASE} 2022 low complexity acoustic scene classification task1,''
  DCASE2022 Challenge, Tech. Rep., 2022.

\bibitem{schmid22KD}
F.~Schmid, S.~Masoudian, K.~Koutini, and G.~Widmer, ``Knowledge distillation
  from transformers for low-complexity acoustic scene classification,'' in
  \emph{DCASE Workshop}, 2022.

\bibitem{koutini22passt}
K.~Koutini, J.~Schl{\"{u}}ter, H.~Eghbal{-}zadeh, and G.~Widmer, ``Efficient
  training of audio transformers with patchout,'' in
  \emph{{Interspeech}}.\hskip 1em plus 0.5em minus 0.4em\relax {ISCA}, 2022.

\bibitem{mixstyle}
J.~Fu, Y.~Zhong, and F.~Yang, ``Adversarial domain generalization with
  mixstyle,'' in \emph{International Conference on Advanced Robotics and
  Mechatronics {(ICARM)}}.\hskip 1em plus 0.5em minus 0.4em\relax {IEEE}, 2022.

\bibitem{kim22rfn}
B.~Kim, S.~Yang, J.~Kim, H.~Park, J.~Lee, and S.~Chang, ``Domain generalization
  with relaxed instance frequency-wise normalization for multi-device acoustic
  scene classification,'' in \emph{{Interspeech}}.\hskip 1em plus 0.5em minus
  0.4em\relax {ISCA}, 2022.

\bibitem{morocutti23dir}
T.~Morocutti, F.~Schmid, K.~Koutini, and G.~Widmer, ``Device-robust acoustic
  scene classification via impulse response augmentation,'' in
  \emph{{EUSIPCO}}.\hskip 1em plus 0.5em minus 0.4em\relax {IEEE}, 2023.

\bibitem{Schmid23tech_report}
F.~Schmid, T.~Morocutti, S.~Masoudian, K.~Koutini, and G.~Widmer, ``{CP-JKU}
  submission to {DCASE23}: Efficient acoustic scene classification with
  cp-mobile,'' DCASE2023 Challenge, Tech. Rep., 2023.

\bibitem{koutini21journal}
K.~Koutini, H.~Eghbal{-}zadeh, and G.~Widmer, ``Receptive field regularization
  techniques for audio classification and tagging with deep convolutional
  neural networks,'' \emph{{IEEE} {ACM} Trans. Audio Speech Lang. Process.},
  2021.

\bibitem{Koutini2019Receptive}
K.~Koutini, H.~Eghbal{-}zadeh, M.~Dorfer, and G.~Widmer, ``The receptive field
  as a regularizer in deep convolutional neural networks for acoustic scene
  classification,'' in \emph{{EUSIPCO}}.\hskip 1em plus 0.5em minus 0.4em\relax
  {IEEE}, 2019.

\bibitem{Sandler18MobileNetsV2}
M.~Sandler, A.~G. Howard, M.~Zhu, A.~Zhmoginov, and L.~Chen, ``Mobilenetv2:
  Inverted residuals and linear bottlenecks,'' in \emph{{CVPR}}.\hskip 1em plus
  0.5em minus 0.4em\relax {IEEE}, 2018.

\bibitem{Howard19MobileNetV3}
A.~Howard, R.~Pang, H.~Adam, Q.~V. Le, M.~Sandler, B.~Chen, W.~Wang, L.~Chen,
  M.~Tan, G.~Chu, V.~Vasudevan, and Y.~Zhu, ``Searching for mobilenetv3,'' in
  \emph{{ICCV}}.\hskip 1em plus 0.5em minus 0.4em\relax {IEEE}, 2019.

\bibitem{Tan19EfficientNet}
M.~Tan and Q.~V. Le, ``Efficientnet: Rethinking model scaling for convolutional
  neural networks,'' in \emph{{ICML}}.\hskip 1em plus 0.5em minus 0.4em\relax
  {PMLR}, 2019.

\bibitem{koutini21dcasesubmission}
K.~Koutini, S.~Jan, and G.~Widmer, ``{CPJKU Submission to DCASE21: Cross-Device
  Audio Scene Classification with Wide Sparse Frequency-Damped {CNNs}},''
  DCASE2021 Challenge, Tech. Rep., 2021.

\bibitem{koutini20dcasesubmission}
K.~Koutini, F.~Henkel, H.~Eghbal-zadeh, and G.~Widmer, ``{{CP-JKU} Submissions
  to {{DCASE}}’20: Low-Complexity Cross-Device Acoustic Scene Classification
  with RF-Regularized {CNNs}},'' DCASE2020 Challenge, Tech. Rep., 2020.

\bibitem{gemmeke17audioset}
J.~F. Gemmeke, D.~P.~W. Ellis, D.~Freedman, A.~Jansen, W.~Lawrence, R.~C.
  Moore, M.~Plakal, and M.~Ritter, ``Audio set: An ontology and human-labeled
  dataset for audio events,'' in \emph{{ICASSP}}.\hskip 1em plus 0.5em minus
  0.4em\relax {IEEE}, 2017.

\bibitem{schmid22cpjku}
F.~Schmid, S.~Masoudian, K.~Koutini, and G.~Widmer, ``{CP-JKU} submission to
  {DCASE22}: Distilling knowledge for low-complexity convolutional neural
  networks from a patchout audio transformer,'' DCASE2022 Challenge, Tech.
  Rep., 2022.

\bibitem{schmid23efficient}
F.~Schmid, K.~Koutini, and G.~Widmer, ``Efficient large-scale audio tagging via
  transformer-to-cnn knowledge distillation,'' in \emph{ICASSP}.\hskip 1em plus
  0.5em minus 0.4em\relax IEEE, 2023.

\end{thebibliography}

%
%
%
%
%
%
%
%
%

\end{sloppy}
\end{document}